\documentclass{aa}  

\usepackage{xfrac}
\usepackage{xcolor}
\usepackage{txfonts}
\usepackage{graphicx}
\usepackage{textcomp}

\usepackage{listings} % For including code

\definecolor{darkbrown}{HTML}{8c4600}
\definecolor{darkblue}{HTML}{1833a1}

\newcommand{\vg}{v_{\rm g}}
\newcommand{\vp}{v_{\rm p}}
\newcommand{\vfrag}{v_{\rm frag}}
\newcommand{\vcoll}{v_{\rm coll}}
\newcommand{\vdrift}{v_{\rm drift}}

\newcommand{\dt}{\Delta t}
\newcommand{\tz}{t_Z}
\newcommand{\tstop}{t_{\rm stop}}
\newcommand{\tgrow}{t_{\rm grow}}

\newcommand{\St}{{\rm St}}
\newcommand{\Stx}{{\rm St}_{\rm x}}

\newcommand{\Stmax}{{\rm St}_{\rm max}}
\newcommand{\Stfrag}{{\rm St}_{\rm frag}}

\newcommand{\cs}{c_{\rm s}}

\newcommand{\Hp}{H_{\rm p}}

\newcommand{\ceff}{\tilde{c}_{\rm s}}
\newcommand{\Heff}{\tilde{H}}
\newcommand{\Omegak}{\Omega_{\rm K}}
\newcommand{\Omegav}{\Omega_{\rm V}}

\newcommand{\Zmax}{Z_{\rm max}}

\newcommand{\rhop}{\rho_{\rm p}}
\newcommand{\rhog}{\rho_{\rm g}}

\begin{document}

\title{Positive Feedback II: How Dust Coagulation inside Vortices Can Form Planetesimals at Low Metallicity}
%\subtitle{}
\titlerunning{Positive Feedback II}

%\fnmsep
\author{
Daniel Carrera\inst{\ref{NMSU}}\thanks{\email{carrera4@nmsu.edu}}
\and Linn E.J. Eriksson\inst{\ref{Stony},\ref{AMNH}}
\and Jeonghoon Lim\inst{\ref{ISU}}
%\and Jeonghoon Lim ({\koreanfont 임정훈})\inst{\ref{ISU}}
\and Wladimir Lyra\inst{\ref{NMSU}}
\and Jacob B. Simon\inst{\ref{ISU}}
}

\institute{
New Mexico State University, Department of Astronomy, PO Box 30001 MSC 4500, Las Cruces, NM 88001, USA\label{NMSU}
\and Department of Physics and Astronomy, Iowa State University, Ames, IA 50010, USA\label{ISU}
\and Institute for Advanced Computational Sciences, Stony Brook University, Stony Brook, NY, 11794-5250, USA\label{Stony}
\and Department of Astrophysics, American Museum of Natural History, 200 Central Park West, New York, NY 10024, USA\label{AMNH}
}

%\date{Received September 30, 20XX}

% \abstract{}{}{}{}{}
% 5 {} token are mandatory
\abstract
{The origin of planetesimals ($\sim$100 km planet building blocks) has confounded astronomers for decades, as numerous growth barriers appear to impede their formation. In a recent paper we proposed a novel interaction where the streaming instability (SI) and dust coagulation work in tandem, with each one changing the environment in a way that benefits the other. This mechanism proved effective at forming planetesimals in the fragmentation-limited inner disk, but much less effective in the drift-limited outer disk, concluding that dust traps may be key to forming planets at wide orbital separations.}
{Here we explore a different hypothesis: That vortices host a feedback loop in which a vortex traps dust, boosting dust coagulation, which in turn boosts vortex trapping.}
{We combine an analytic model of vortex trapping with an analytic model of fragmentation limited grain growth that accounts for how dust concentration dampens gas turbulence.}
{We find a powerful synergy between vortex trapping and dust growth. For $\alpha \le 10^{-3}$ and solar-like metallicity this feedback loop consistently takes the grain size and dust density into the planetesimal formation region of the streaming instability (SI). Only in the regime of strong turbulence ($\alpha \ge 3\times 10^{-3}$) does the system often converge to a steady state below the SI criterion.}
{The combination of vortex trapping with dust coagulation is an even more powerful mechanism than the one involving the SI. It is effective at lower metallicity and across the whole disk --- anywhere that vortices form.}

%% I don't know what keywords A&A uses, but AAS Journals
%% use Unified Astronomy Thesaurus concepts:

\maketitle

\nolinenumbers

%%%%%%%%%%%%%%%%%%%%%%%%%%%%%%%%%%%%%%%%%%%%%%%%%%%%%%%%%%%%
% INTRODUCTION
%%%%%%%%%%%%%%%%%%%%%%%%%%%%%%%%%%%%%%%%%%%%%%%%%%%%%%%%%%%%
\section{Introduction}
\label{sec:intro}

Despite decades of research, we still lack a coherent picture of planet formation. When stars are young, they are surrounded by a circumstellar disk of gas and dust. The dust component must give rise to planetary building blocks such as planetesimals and embryos that can later become rocky planets or the cores of giant planets. The current open question regarding the origin of these building blocks is as fundamental as can be: What is the mechanism that converts dust grains into $>$km-sized bodies?

Once the disk is established, collisions between micron-sized grains leads to rapid growth until the grains reach $\sim$mm-cm sizes, at which point they encounter two main barriers:

\begin{itemize}
\item Fragmentation Barrier: As grains grow in size, their collision speed increases \citep{Ormel_2007} until it overcomes the material strength of the grains \citep[e.g.,][]{Guttler_2010}.

\item Radial Drift Barrier: As grains grow, aerodynamic drag makes them drift toward the star with increasing speed \citep{Weidenschilling_77} until the drift timescale is shorter than the grain growth timescale \citep{Birnstiel_2012}.
\end{itemize}

Current efforts to overcome these barriers generally focus on aerodynamic processes that concentrate dust: If the dust density is sufficiently high, the collective gravity of dust grains can lead to gravitational collapse, giving rise to large $>$km-sized bodies, leapfrogging the intermediate sizes. Two prominent mechanisms include the Streaming Instability (SI), which collects dust into  filaments \citep{Youdin_2005,Johansen_2007a}, and vortices which trap dust \citep{Barge_1995,Adams_1995,Tanga_1996}. Both mechanisms have been shown to produce self-gravitating dust clumps \citep[e.g.,][resp]{Johansen_2007a,Lyra_2024}, but both have important limitations.

For a solar-like dust-to-gas ratio of $Z = 0.01$, dust growth models predict dust sizes of $\St \sim 0.01$ \citep{Drazkowska_2021}, where $\St = \tstop\Omegak$ is the stopping time $\tstop$ normalized by the Keplerian frequency $\Omegak$. But the SI requires high $(Z,\St)$ to work \citep[e.g.,][]{Carrera_2015,Lim_2024a} and it has not been shown to form planetesimals for $Z = 0.01, \St = 0.01$.  Conversely, vortex trapping struggles for $\St = 0.03$, $Z = 0.01$ and $\alpha = 3\times 10^{-4}$ \citep{Lyra_2024} where $\alpha$ is the turbulence parameter \citep{Shakura_1973}. It has not been able to work for realistic $\alpha \sim 10^{-3} - 10^{-2}$, \citep{Lesur_2010,Lyra_2011}, $Z = 0.01$, and $\St = 0.01$.

Furthermore, there are many known exoplanets around sub-solar metallicity stars (e.g., GJ 9827c, and Kepler 37d \& 408b are $0.2 - 2 M_\oplus$ planets around stars with $0.003\le Z_\star \le 0.006$), so any planetesimal formation model must work for $Z < 0.01$. Here we find a mechanism that bypasses these problems by simultaneously increasing dust concentration and $\St$ while decreasing turbulence.

Recently we proposed a mechanism where the SI and dust growth work in tandem, forming a feedback loop where each process enhances the other \citep[][``Paper I'']{Carrera_2025}: The SI concentrates dust, which dampens turbulence \citep{Johansen_2009} and slows radial drift, promoting grain growth. In turn, grain growth makes the SI more effective. This feedback proved extremely effective in the fragmentation-limited regime, taking the system straight toward the region where the SI is thought to produce planetesimals. But in the drift-limited regime, the gains were modest.

Here we present a follow-up investigation with a different mix of mechanisms: We propose that vortex trapping also exhibits a feedback loop with dust growth. Since vortices also collect particles, they also dampen turbulence, which promotes grain growth. At the same time, larger grains (up to $\St \lesssim 1$) concentrate more strongly. Because vortices are true dust traps, this mechanism completely eliminates the radial drift barrier, making it especially important in the drift-limited outer disk.

This paper is organized as follows. We present our model in section \S\ref{sec:model}. We describe how mass loading affects the fragmentation barrier, and how we combine grain growth and vortices into a feedback loop. Section \S\ref{sec:results} shows our final results. We discuss in section \S\ref{sec:discussion} and draw conclusions in section \S\ref{sec:conclusion}.

%%%%%%%%%%%%%%%%%%%%%%%%%%%%%%%%%%%%%%%%%%%%%%%%%%%%%%%%%%%%
% ANALYTIC MODEL
%%%%%%%%%%%%%%%%%%%%%%%%%%%%%%%%%%%%%%%%%%%%%%%%%%%%%%%%%%%%
\section{Model}
\label{sec:model}

We use the same model for turbulence dampening as Paper I. We include a summary in Appendix \ref{appendix:model:Stfrag}.

%--------------------------------------------------%
% MASS LOADING AND VORTEX CONCENTRATION
%--------------------------------------------------%
\subsection{Vortex Trapping}
\label{sec:model:vortex}

\begin{figure}[t]
\centering
\includegraphics[width=0.49\textwidth]{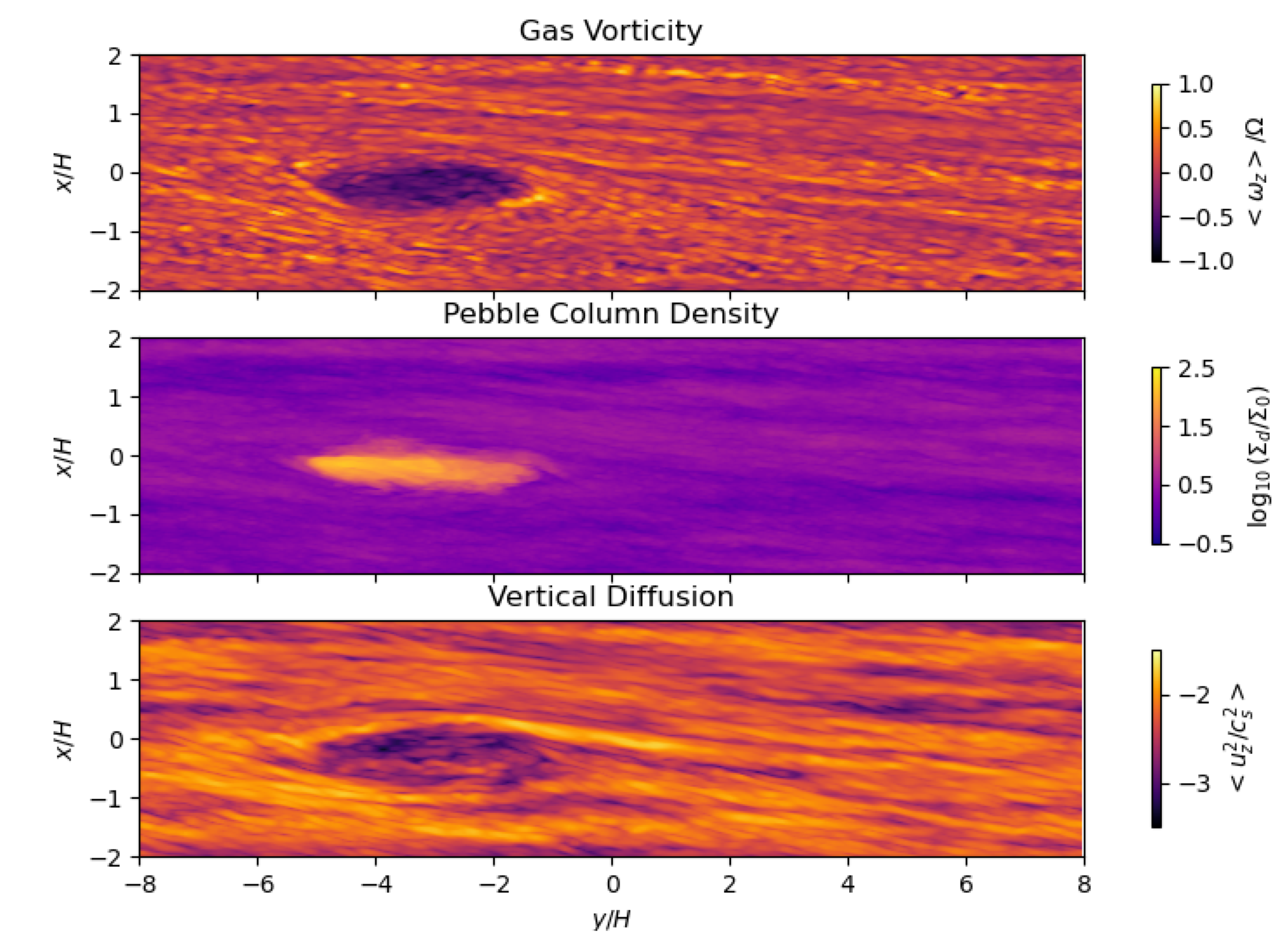}
\caption{Gas vorticity (top), pebble column density (middle), and vertical diffusion (bottom) for the $512^3$ simulation of \citet{Lyra_2024}.
\label{fig:vortex}}
\end{figure}

Inside a vortex the dust density follows a Gaussian profile with constant density along ellipses of equal aspect ratio \citep{Lyra_2013}. The Gaussian peaks at the center of the vortex, reaching a maximum column dust-to-gas ratio of

\begin{equation}\label{eqn:Zmax}
    \Zmax = Z_{\rm disk} \left(1 + \frac{\St}{\alpha}\right)
\end{equation}

\noindent
Figure \ref{fig:vortex} shows the vorticity, pebble column density, and vertical diffusion inside a vortex using the recent simulation data from \citet{Lyra_2024}. Notice that the vortex has a distinct $\alpha$, independent from the rest of the disk. Vortices produce their own turbulence via the elliptic instability \citep{Lesur_2010,Lyra_2011}. Notice also the dust concentration inside the vortex with a peak near the center. \citet{Lyra_2024} showed that their runs have maximum dust density consistent with Equation \ref{eqn:Zmax}, at least up to the point where gravitational instability leads to collapse. To estimate the vortex trapping timescale, we start with the drift velocity of solid grains relative to the gas inside a vortex \citep{Lyra_2013}

\begin{equation}
    \vec{v}_{\rm drift} = \tstop \vec{\nabla h}.
\end{equation}

\noindent
where $h$ is enthalpy. We refer to \citet{Lyra_2013} for the full expression of $\vec{\nabla h}$ inside the vortex, but the key result is that, for a typical vortex aspect ratio of 4, $H\Omegak^2 \le |\vec{\nabla h}| \le 4H \Omegak^2$ at the boundary, decreasing toward zero at the center. Note also that, in general, $\vec{\nabla h}$ does not point exclusively toward the center of the vortex, so that the ``radial'' speed is $v_{\rm rad} \le \vdrift$\footnote{We use the term ``radial'' to mean ``toward the center of the ellipse''.}. Nonetheless, these constraints provide a useful ballpark estimate of the radial drift rate of solid grains

\begin{equation}
    v_{\rm rad} \lesssim \tstop\,\Omegak^2\,H = \St\,\cs
\end{equation}

\noindent
Let $\tz$ be the vortex trapping timescale. Again, for a vortex with an aspect ratio of 4, the radial distance traversed by the grains is between $H$ and $4H$, so that

\begin{equation}\label{eqn:tZ}
    \tz \gtrsim \frac{H}{v_{\rm rad}}
        \gtrsim \frac{H}{\St\,\cs}
        = \frac{1}{\St\,\Omegak}.
\end{equation}

\noindent
Comparing this expression against the simulations of \citet{Lyra_2024}, we find support for the $\tz \propto 1/\St$ scaling, and indeed we find that $\tz$ appears to be in the vicinity of $\tz \sim 4(\St\,\Omegak)^{-1}$. To cover the range of uncertainty in $\tz$, we run our semi-analytic model twice, spanning an order of magnitude in $\tz$

\begin{equation}
    \tz \in \left[
        \frac{1}{\St\,\Omegak} ,
        \frac{10}{\St\,\Omegak}
    \right]
\end{equation}

\noindent
It is worth noting that vortex trapping cannot continue indefinitely. When $\St \gg 1$, solid grains can no longer be trapped and escape the vortex \citep[e.g.,][]{Raettig_2015}. In practice, we only explore the parameter space for $\St \le 0.1$ because that is the most relevant region for determining whether the local conditions are consistent with planetesimal formation via the SI.

%--------------------------------------------------%
% VERTICAL SEDIIMENTATION
%--------------------------------------------------%
\subsection{Vertical Sedimentation}
\label{sec:model:sedimentation}

Next, we  add an expression for dust sedimentation that accounts for mass loading. A common approach is to model $\rhop(z)$ as a Gaussian profile with scale height $\Hp = H\sqrt{\alpha/(\St + \alpha)}$ so that the midplane dust-to-gas ratio is given by $\epsilon = Z (H/\Hp)$ \citep{Youdin_2007}. However, this expression does not account for mass loading. Using the colloid approximation, \citet{Yang_2020} defined the effective scale height of the dust-gas mixture $\Heff \equiv \ceff/\Omega$, and \citet{Lim_2024a} showed that

\begin{equation}\label{eqn:Hp}
    \Hp = \Heff \sqrt{
        \left(\frac{\Pi}{5}\right)^2 +
        \frac{\alpha}{\alpha + \St}
    }
\end{equation}

\noindent
is a better predictor of the dust scale height. The $(\Pi/5)^2$ term estimates the amount of turbulence caused by the SI. In this work we make two further modifications: We assume that inside the vortex the SI is not the dominant source of turbulence so that the $(\Pi/5)^2$ term can be neglected, and we use Equation \ref{eqn:ceff} for $\ceff$ instead of the colloid approximation. The two expressions agree for $\St \ll 1$. This gives us an expression for the midplane dust-to-gas ratio that is valid even when neither $\St$ nor $\epsilon$ are negligible.

\begin{equation}\label{eqn:epsilon_gaussian}
    \epsilon
    = Z\frac{H}{\Hp}
    = Z \sqrt{\frac{1 + \St + \epsilon}{1 + \St}}
        \sqrt{1 + \frac{\St}{\alpha}}
\end{equation}

\noindent
Equation \ref{eqn:epsilon_gaussian} assumes that the dust has a Gaussian profile. The true profile is more centrally peaked \citep{Lim_2024a}, meaning that Equation \ref{eqn:epsilon_gaussian} is a conservative estimate. This is a quadratic on $\epsilon$. Let $\zeta \equiv Z^2 (1+\St/\alpha) / (1+\St)$ and solve
 
\begin{equation}\label{eqn:epsilon}
    \epsilon = \frac{\zeta + \sqrt{\zeta^2 + 4\zeta(1+\St)}}{2}
\end{equation}

\noindent
Notice that we did not include Equation \ref{eqn:Zmax} in this expression. That would be a good option if we were to treat $\St$ as a static quantity. But we are interested $(\St,Z)$ as dynamic quantities that evolve together and respond to one another. As a result, the expressions for $\Stfrag$ (Equation \ref{eqn:Stfrag}) and $\Zmax$ (Equation \ref{eqn:Zmax}) are dynamic targets that the system is steadily moving toward. This is described in more detail in the next section.

%--------------------------------------------------%
% FEEDBACK LOOP
%--------------------------------------------------%
\subsection{Feedback Loop}
\label{sec:model:feedback}

The feedback loop arises from the co-evolution of dust growth, vortex trapping, and sedimentation, as each of these processes changes the environment for the others. The model parameters are the disk's column dust to gas ratio $Z_{\rm disk} = 0.01$, the turbulence parameter $\alpha \in [10^{-4},3 \times 10^{-3}]$, and the classic fragmentation barrier $\Stx \in [0.01,0.04]$. Using $\Stx$ as an input parameter allows us to explore the problem without assuming a particular disk model.

Recall that our objective is to from planetesimals in the drift-dominated outer disk. Therefore, dust grains enter the vortex with a small, drift-limited, grain size $\St_0 < \Stx$, and then grow inside the vortex. For the sake of simplicity, we set $\St_0 \equiv 0.1\Stx$, initialize $\epsilon$ with Equation \ref{eqn:epsilon}, and apply the following algorithm:

\begin{enumerate}
    \item Update $\Stfrag,\tgrow,\Zmax,\tz$ (Equations \ref{eqn:Stfrag}, \ref{eqn:tgrow}, \ref{eqn:Zmax}, \ref{eqn:tZ}).
    \item Let $\dt = 0.1\min(\tz,\tgrow)$ be the iteration timestep.
    \item Update $(\St,Z)$ at the same time
    \begin{itemize}
        \item $\St = \min\left[
                        \Stmax,
                        \St\cdot\exp(\dt/\tgrow)
                    \right]$
        \item $Z = \min\left[
                        \Zmax, Z\cdot\exp(\dt/\tz)
                    \right]$
    \end{itemize}
    \item Update $\epsilon$ (Equation \ref{eqn:epsilon}).
\end{enumerate}

%%%%%%%%%%%%%%%%%%%%%%%%%%%%%%%%%%%%%%%%%%%%%%%%%%%%%%%%%%%%
% RESULTS
%%%%%%%%%%%%%%%%%%%%%%%%%%%%%%%%%%%%%%%%%%%%%%%%%%%%%%%%%%%%

\begin{figure*}[!ht]
\centering
\includegraphics[width=0.9\textwidth]{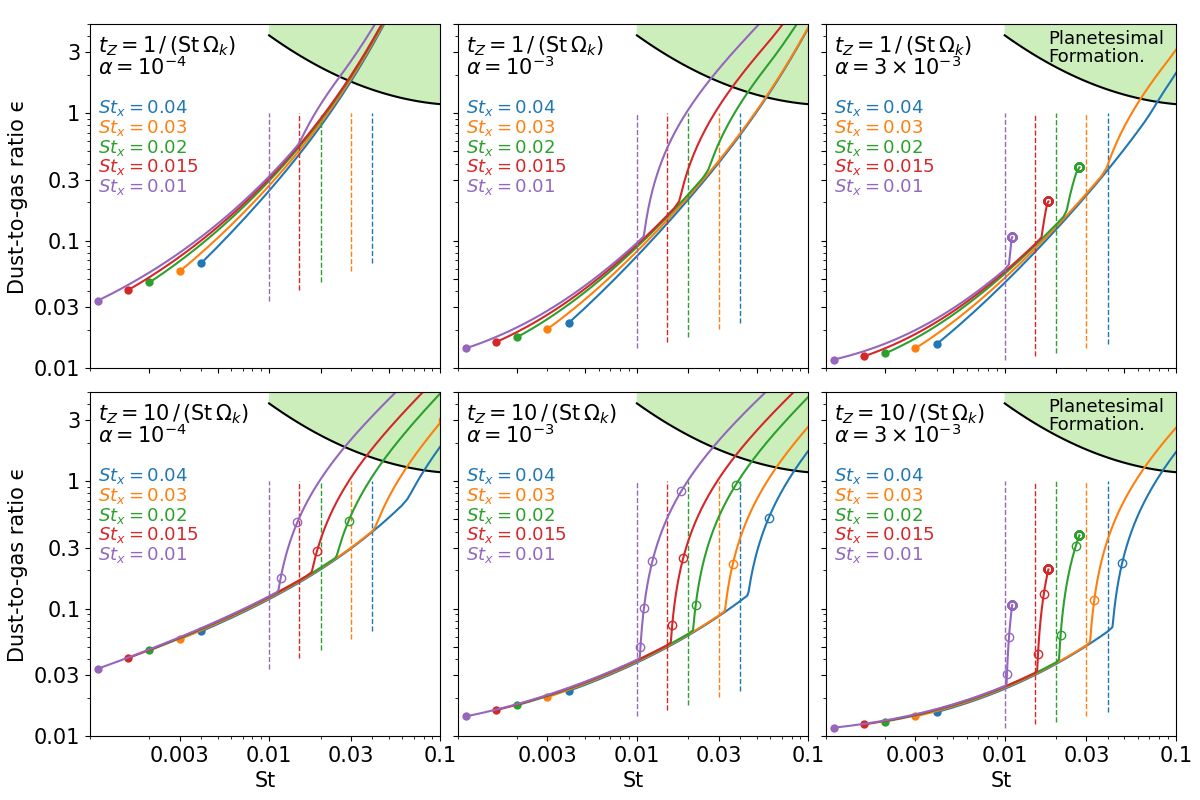}
\caption{Growth tracks for a range of grain sizes $\St$, levels of turbulence $\alpha$, and vortex trapping timescales $\tz$. Every run has $Z = 0.01$. The vertical dashed lines are $\Stx$, but the growth tracks start at $\St_0 = 0.1\Stx$, as a proxy for the small, drift-limited grains entering the vortex. Solid circles mark the start of the growth tracks and open circles mark every 100 orbits to show the time evolution. The green region is the SI planetesimal formation \citep{Lim_2024a}. Most scenarios result in growth tracks that reach this region. However, for sufficiently large $\alpha$ and small $\Stx$, the growth tracks converge to a steady state outside the planetesimal formation region (the thick circles are multiple iterations plotted on top of each other).
\label{fig:particle_tracks}}
\end{figure*}

\section{Results}
\label{sec:results}

Figure \ref{fig:particle_tracks} shows the evolution of $(\St,\epsilon)$ at the center of the vortex for twenty-four simulations. We explore a range of turbulence values $10^{-4} \le \alpha \le 3\times 10^{-3}$, and vortex trapping timescales $1 \le \tz(\St\Omegak)\le 10$. Every simulation has $Z = 0.01$. We are mainly interested in vortices in the outer disk, where dust grains are limited by radial drift instead of fragmentation. That means that dust grains might enter the vortex well below the fragmentation limit. To capture this, we set the initial grain size to $\St_0 = 0.1\Stx$, where $\Stx$ is the value of the fragmentation limit most often encountered in the literature (Equation \ref{eqn:Stx}). We treat $\Stx$ as an input parameter so that our analysis remains agnostic to the disk model.

First and foremost, we find that this mechanism is extremely effective. Nearly every scenario leads to $(\St,\epsilon)$ evolution tracks that point directly into the planetesimal formation region for the SI \citep[green region;][]{Lim_2024a}. We find that ``\textit{vortex trapping + dust growth}'' is a far more powerful mechanism than the ``\textit{SI + dust growth}'' that we explored in Paper I. The mechanism in Paper I was ineffective in the drift-limited regime and required $Z > 0.01$ in the fragmentation-limited regime. Replacing the SI with vortices allows us to run all models with $Z = 0.01$. In fact, a companion work \citep{Eriksson_2025} shows that this mechanism can form planetesimals in the ultra-low metallicity disks of the early universe ($Z \ge 0.0004$).

Secondly, we do find that this mechanism can stall for strong turbulence relative to the grain size ($\alpha = 3\times 10^{-3}, \Stx \le 0.02$ in Figure \ref{fig:particle_tracks}). It is worth noting that vortices can have lower turbulence than their surroundings. Figure \ref{fig:vortex} shows an example vortex where the surroundings have $\alpha \approx 10^{-2}$ but the interior has only $\alpha \approx 10^{-3}$. This occurs because the vortex has its own $\alpha$, driven by the elliptic instability, which can be quite different from the $\alpha$ value in the rest of the disk \citep{Lesur_2010,Lyra_2011}.

%%%%%%%%%%%%%%%%%%%%%%%%%%%%%%%%%%%%%%%%%%%%%%%%%%%%%%%%%%%%%%
\section{Discussion}
\label{sec:discussion}

%--------------------------------------------------%
% BOUNCING BARRIER
%--------------------------------------------------%
\subsection{Bouncing Barrier vs Dust Traps}
\label{sec:discussion:bouncing}

Perhaps the most important caveat for this investigation is that grain sizes might be limited by ``bouncing'' rather than fragmentation. That is to say, grain growth stalls because grain collisions result in bouncing instead of sticking \citep{Zsom_2010}.

One of the most important properties of this barrier is that, if present, it can lead to grains that are much smaller than those of the fragmentation limit \citep{Zsom_2010}. However, it is not clear that this is the case. Whether collisions lead to sticking or bouncing depends on the detailed properties of the grains, such as their shape, surface tension, and porosity. Furthermore, \citet{Jungmann_2021} have made a strong case that the bouncing barrier may be overcome by electrostatic forces.

Suppose that the bouncing barrier is present. The bouncing barrier resembles fragmentation in that they are both a limit on the collision speed between grains $\vcoll = \cs \sqrt{3\alpha\St}$. The critical difference is that bounce-limited grains are around an order of magnitude smaller \citep{Dominik_2024}. That means that the $(\St,\epsilon)$ evolution tracks should have a similar shape to those in Figure \ref{fig:particle_tracks}, but will require either higher $Z$ or lower $\alpha$ to reach the planetesimal formation region.

This leads us to an important process that we omitted: pebble flux. We kept the total dust mass inside the vortex constant. In a real disk there is a steady influx of dust grains drifting from the outer disk, which are captured by the vortex. Therefore, $Z$ grows over the lifetime of the vortex. This might be the key to overcoming the bouncing barrier if it is present.

%%%%%%%%%%%%%%%%%%%%%%%%%%%%%%%%%%%%%%%%%%%%%%%%%%%%%%%%%%%%%%
\section{Conclusions}
\label{sec:conclusion}

We present a novel mechanism where dust growth and vortex trapping work in tandem, as each one changes the environment in a way that enhances the other: Vortices eliminate the radial drift barrier and concentrate grains, which dampens turbulence. Lower turbulence allows fragmentation-limited grains to grow, and larger grains concentrate more strongly. This new interaction is potentially more powerful than the one involving the SI that we reported in Paper I: 

\begin{enumerate}
\item Unlike the mechanism of Paper I, the one presented here is effective for $Z = 0.01$ and $\St = 0.01$  (Figure \ref{fig:particle_tracks}), making it fully compatible with dust evolution models.

\item The mechanism presented here is active wherever vortices form. Crucially, it is active in the drift-dominated outer disk, where the mechanism of Paper I is not effective.
\end{enumerate}

\noindent
We did find that, for sufficiently high turbulence ($\alpha \ge 3\times 10^{-3}$; higher than suggested by vortex simulations) the system can stall below the planetesimal formation threshold. However, in a real disk this would be mitigated by the fact that vortices continuously trap grains as they drift from the outer disk. The total dust mass in the vortex increases for as long as the vortex lives.

Altogether, the combination of vortices and dust growth in a feedback loop appears to bridge the gap between the dust growth barriers and planetesimal formation mechanisms. This novel mechanism works entirely within the $(\St,Z)$ constraints predicted by dust evolution models, and it is effective anywhere that vortices form.

%%%%%%%%%%%%%%%%%%%%%%%%%%%%%%%%%%%%%%%%%%%%%%%%%%%%%%%%%%%%%%
\begin{acknowledgements}
DC, WL, and JBS acknowledge support from NASA under {\em Emerging Worlds} through grant 80NSSC25K7414. J.L. acknowledges support from NASA under the Future Investigators in NASA Earth and Space Science and Technology grant 80NSSC22K1322. LE acknowledges the support from NASA via the Emerging Worlds program (80NSSC25K7117), as well as the Institute for Advanced Computational Science Postdoctoral Fellowship.
\end{acknowledgements}


\bibliographystyle{aa}
\bibliography{references}

\begin{thebibliography}{34}
\expandafter\ifx\csname natexlab\endcsname\relax\def\natexlab#1{#1}\fi

\bibitem[{{Adams} \& {Watkins}(1995)}]{Adams_1995}
{Adams}, F.~C. \& {Watkins}, R. 1995, \apj, 451, 314

\bibitem[{{Barge} \& {Sommeria}(1995)}]{Barge_1995}
{Barge}, P. \& {Sommeria}, J. 1995, \aap, 295, L1

\bibitem[{{Birnstiel} {et~al.}(2012){Birnstiel}, {Klahr}, \& {Ercolano}}]{Birnstiel_2012}
{Birnstiel}, T., {Klahr}, H., \& {Ercolano}, B. 2012, \aap, 539, A148

\bibitem[{{Carrera} {et~al.}(2015){Carrera}, {Johansen}, \& {Davies}}]{Carrera_2015}
{Carrera}, D., {Johansen}, A., \& {Davies}, M.~B. 2015, \aap, 579, A43

\bibitem[{{Carrera} {et~al.}(2025){Carrera}, {Lim}, {Eriksson}, {Lyra}, \& {Simon}}]{Carrera_2025}
{Carrera}, D., {Lim}, J., {Eriksson}, L. E.~J., {Lyra}, W., \& {Simon}, J.~B. 2025, arXiv e-prints, arXiv:2503.03105

\bibitem[{{Chang} \& {Oishi}(2010)}]{Chang_2010}
{Chang}, P. \& {Oishi}, J.~S. 2010, \apj, 721, 1593

\bibitem[{{Chen} \& {Lin}(2018)}]{Chen_2018}
{Chen}, J.-W. \& {Lin}, M.-K. 2018, \mnras, 478, 2737

\bibitem[{{Cuzzi} {et~al.}(1993){Cuzzi}, {Dobrovolskis}, \& {Champney}}]{Cuzzi_1993}
{Cuzzi}, J.~N., {Dobrovolskis}, A.~R., \& {Champney}, J.~M. 1993, \icarus, 106, 102

\bibitem[{{Dominik} \& {Dullemond}(2024)}]{Dominik_2024}
{Dominik}, C. \& {Dullemond}, C.~P. 2024, \aap, 682, A144

\bibitem[{{Drazkowska} {et~al.}(2021){Drazkowska}, {Stammler}, \& {Birnstiel}}]{Drazkowska_2021}
{Drazkowska}, J., {Stammler}, S.~M., \& {Birnstiel}, T. 2021, \aap, 647, A15

\bibitem[{{Eriksson} {et~al.}(2025){Eriksson}, {Menon}, {Carrera}, {Lyra}, \& {Burkhart}}]{Eriksson_2025}
{Eriksson}, L. E.~J., {Menon}, S., {Carrera}, D., {Lyra}, W., \& {Burkhart}, B. 2025, arXiv e-prints, arXiv:2503.11877

\bibitem[{{G{\"u}ttler} {et~al.}(2010){G{\"u}ttler}, {Blum}, {Zsom}, {Ormel}, \& {Dullemond}}]{Guttler_2010}
{G{\"u}ttler}, C., {Blum}, J., {Zsom}, A., {Ormel}, C.~W., \& {Dullemond}, C.~P. 2010, \aap, 513, A56

\bibitem[{{Johansen} {et~al.}(2007){Johansen}, {Oishi}, {Mac Low}, {Klahr}, {Henning}, \& {Youdin}}]{Johansen_2007a}
{Johansen}, A., {Oishi}, J.~S., {Mac Low}, M.-M., {et~al.} 2007, \nat, 448, 1022

\bibitem[{{Johansen} {et~al.}(2009){Johansen}, {Youdin}, \& {Mac Low}}]{Johansen_2009}
{Johansen}, A., {Youdin}, A., \& {Mac Low}, M.-M. 2009, \apjl, 704, L75

\bibitem[{{Jungmann} \& {Wurm}(2021)}]{Jungmann_2021}
{Jungmann}, F. \& {Wurm}, G. 2021, \aap, 650, A77

\bibitem[{{Laibe} \& {Price}(2014)}]{Laibe_2014}
{Laibe}, G. \& {Price}, D.~J. 2014, \mnras, 440, 2136

\bibitem[{{Lesur} \& {Papaloizou}(2010)}]{Lesur_2010}
{Lesur}, G. \& {Papaloizou}, J.~C.~B. 2010, \aap, 513, A60

\bibitem[{{Lim} {et~al.}(2024){Lim}, {Simon}, {Li}, {Armitage}, {Carrera}, {Lyra}, {Rea}, {Yang}, \& {Youdin}}]{Lim_2024a}
{Lim}, J., {Simon}, J.~B., {Li}, R., {et~al.} 2024, \apj, 969, 130

\bibitem[{{Lin} \& {Youdin}(2017)}]{Lin_2017}
{Lin}, M.-K. \& {Youdin}, A.~N. 2017, \apj, 849, 129

\bibitem[{{Lyra} \& {Klahr}(2011)}]{Lyra_2011}
{Lyra}, W. \& {Klahr}, H. 2011, \aap, 527, A138

\bibitem[{{Lyra} \& {Lin}(2013)}]{Lyra_2013}
{Lyra}, W. \& {Lin}, M.-K. 2013, \apj, 775, 17

\bibitem[{{Lyra} {et~al.}(2024){Lyra}, {Yang}, {Simon}, {Umurhan}, \& {Youdin}}]{Lyra_2024}
{Lyra}, W., {Yang}, C.-C., {Simon}, J.~B., {Umurhan}, O.~M., \& {Youdin}, A.~N. 2024, \apjl, 970, L19

\bibitem[{{Ormel} \& {Cuzzi}(2007)}]{Ormel_2007}
{Ormel}, C.~W. \& {Cuzzi}, J.~N. 2007, \aap, 466, 413

\bibitem[{{Raettig} {et~al.}(2015){Raettig}, {Klahr}, \& {Lyra}}]{Raettig_2015}
{Raettig}, N., {Klahr}, H., \& {Lyra}, W. 2015, \apj, 804, 35

\bibitem[{{Schr{\"a}pler} \& {Henning}(2004)}]{Schraepler_2004}
{Schr{\"a}pler}, R. \& {Henning}, T. 2004, \apj, 614, 960

\bibitem[{{Shakura} \& {Sunyaev}(1973)}]{Shakura_1973}
{Shakura}, N.~I. \& {Sunyaev}, R.~A. 1973, \aap, 24, 337

\bibitem[{{Shi} \& {Chiang}(2013)}]{Shi_2013}
{Shi}, J.-M. \& {Chiang}, E. 2013, \apj, 764, 20

\bibitem[{{Tanga} {et~al.}(1996){Tanga}, {Babiano}, {Dubrulle}, \& {Provenzale}}]{Tanga_1996}
{Tanga}, P., {Babiano}, A., {Dubrulle}, B., \& {Provenzale}, A. 1996, \icarus, 121, 158

\bibitem[{{Voelk} {et~al.}(1980){Voelk}, {Jones}, {Morfill}, \& {Roeser}}]{Voelk_1980}
{Voelk}, H.~J., {Jones}, F.~C., {Morfill}, G.~E., \& {Roeser}, S. 1980, \aap, 85, 316

\bibitem[{{Weidenschilling}(1977)}]{Weidenschilling_77}
{Weidenschilling}, S.~J. 1977, \mnras, 180, 57

\bibitem[{{Yang} \& {Zhu}(2020)}]{Yang_2020}
{Yang}, C.-C. \& {Zhu}, Z. 2020, \mnras, 491, 4702

\bibitem[{{Youdin} \& {Goodman}(2005)}]{Youdin_2005}
{Youdin}, A.~N. \& {Goodman}, J. 2005, \apj, 620, 459

\bibitem[{{Youdin} \& {Lithwick}(2007)}]{Youdin_2007}
{Youdin}, A.~N. \& {Lithwick}, Y. 2007, \icarus, 192, 588

\bibitem[{{Zsom} {et~al.}(2010){Zsom}, {Ormel}, {G{\"u}ttler}, {Blum}, \& {Dullemond}}]{Zsom_2010}
{Zsom}, A., {Ormel}, C.~W., {G{\"u}ttler}, C., {Blum}, J., \& {Dullemond}, C.~P. 2010, \aap, 513, A57

\end{thebibliography}

%%%%%%%%%%%%%%%%%%%%%%%%%%%%%%%%%%%%%%%%%%%%%%%%%%

%%%%%%%%%%%%%%%%% APPENDICES %%%%%%%%%%%%%%%%%%%%%
\appendix

%--------------------------------------------------%
% MASS LOADING AND THE FRAGMENTATION BARRIER
%--------------------------------------------------%
\section{Mass Loading and the Fragmentation Barrier}
\label{appendix:model:Stfrag}

We use the same model of turbulence dampening as Paper I, and we refer to that paper for details. What follows is a short summary of the model.

The most common way to model mass loading is to treat the gas-dust mixture as a colloidal suspension where dust contributes to the inertial of the fluid but not to its pressure \citep{Chang_2010,Shi_2013,Laibe_2014,Lin_2017,Chen_2018}. This approach is valid in the limit as $\St \rightarrow 0$, but for our investigation we are interested in the case where $\St$ is not negligible, so that the fluid might not be well approximated by a colloid. In Paper I we approach the problem from the point of view of energy conservation, where there is a finite energy source that has to be partitioned between the gas and dust components

\begin{equation}\label{eqn:turbulent_energy}
    E = \frac{1}{2} \rhog \vg^2
      + \frac{1}{2} \rhop \vp^2
      = \frac{1}{2} \rhog \vg^2
        \left( 1 + \epsilon\frac{\vp^2}{\vg^2} \right)
      = \text{Const}
\end{equation}

\noindent
where $\vg$ and $\vp$ are the root-mean-squared velocities of the gas and dust and $\epsilon$ is the dust-to-gas ratio. Using the fact that $\vp = \vg / \sqrt{1 + \St}$ \citep{Voelk_1980,Cuzzi_1993,Schraepler_2004}, Paper I showed that the gas velocity can be expressed as

\begin{eqnarray}
    \vg &=& \sqrt{\alpha}\ceff
    \label{eqn:Vgas}
    \\
    \ceff &\equiv& \cs \sqrt{\frac{1 + \St}{1 + \St + \epsilon}}
    \label{eqn:ceff}
\end{eqnarray}

\noindent
where $\ceff$ is the ``effective'' sound speed of the gas-dust mixture. In the limit as $\St \rightarrow 0$, Equations \ref{eqn:Vgas} and \ref{eqn:ceff} reduce to the colloid approximation.

The fragmentation barrier occurs when the collision speed between grains $\vcoll = \vg\sqrt{3\St}$ \citep{Ormel_2007} reaches the fragmentation speed $\vfrag$ of the grain material. Combined with Equation \ref{eqn:Vgas} we obtain

\begin{eqnarray}
    \Stfrag
    &=& \Stx \left(1 + \frac{\epsilon}{1 + \St} \right)
        \label{eqn:Stfrag}\\
    \St_{\rm x} &\equiv& \frac{\vfrag^2}{3\alpha\cs^2}\label{eqn:Stx}
\end{eqnarray}

\noindent
where $\Stx$ is the usual definition of $\Stfrag$ commonly found in the literature. In other words, mass loading boosts the fragmentation barrier by a factor of $1 + \epsilon/(1 + \St)$. \citet{Birnstiel_2012} derive the grain growth rate $\tgrow = 1/(Z\Omegak)$. For dust growth inside a vortex we will replace this with

\begin{equation}\label{eqn:tgrow}
    \tgrow = \frac{1}{Z\Omegav}.
\end{equation}

\noindent
where $\Omegav$ is the vortex frequency. In practice, $\Omegav \approx \Omegak$, and we adopt the typical value of $\Omegav = 0.5\Omegak$ \citep{Lyra_2013}.

\end{document}